%% file: paper.tex
\documentclass{jfp}
\usepackage{amssymb}

\title{Blind graph rewriting systems}
\author{Anton Salikhmetov}
\email{anton.salikhmetov@gmail.com}

\newcommand{\Z}{{\mathbb Z}}

\begin{document}
\maketitle[t]

\begin{abstract}
We consider a simple (probably, the simplest) structure for
random access memory.
This structure can be used to construct a universal system with
nearly void processor, namely, we demonstrate that the processor
of such a system may have empty instruction set, in a more strong
manner than the existing ZISC (zero instruction set computer
based on ideas for artificial neural networks) and NISC
architecture (no instruction set computing).
More precisely, the processor will be forbidden to analyze
any information stored in the memory, the latter being the only
state of such a machine.
This particular paper is to cover an isolated aspect of the idea,
specifically, to provide the logical operations embedded into
a system without any built-in conditional statements.
\end{abstract}

\section{Graph rewriting systems}

Graph rewriting systems appeared to be essential when Wadsworth
used sharing of $\lambda$-expressions, practically inventing
what is nowadays called lazy evaluation~\cite{wadsworth}.
One particular branch of further developments based on this idea,
including ``call-by-need'' evaluation strategy for functional
programming languages, was focused on optimal reduction, that is
a reduction mechanism that uses optimal sharing to minimize
the reduction steps to achieve normal form if any.
Asymptotically by complexity of $\lambda$-expressions,
optimal reduction is the best possible evaluation technique.

\input inet

The rest of this paper is an attempt to find the simplest
automata able to implement arbitrary graph rewriting system.
As we will see, there exist automata with static transition
function which in some sense does not rely on the current state.
We believe this kind of evaluation could result in interesting
forms of computation, for instance, based on RAM with CPU that
performs one and the same chain of move instructions.

\input op
\input embed
\input app

\input biblio
\end{document}

%% file: inet.tex
The first algorithm for optimal reduction was that by Lamping
who formulated his results in a very special form of
graph rewriting system~\cite{lamping}.
Specifically, his system had such properties as strong
confluence and locality, the latter being useful for pattern
matching and tracking of redexes.

The idea behind graph rewriting systems similar to that by
Lamping was caught by Lafont who generalized and described them
as interaction systems~\cite{lafont90}.
The latter consist of a signature, that is, a set of agents
that constitute interaction nets and rules for interaction
between agents.
Lafont introduced a simple language for interaction systems, and
the Lamping algorithm can be defined using this language as well:
\begin{eqnarray*}
\Sigma = &\{(\epsilon, 0)\} \cup \{(\triangle_i, 2) | i \in \Z\}
\cup \{(\sqcap_i, 1) | i \in \Z\} \cup
\{(\cap_i, 1) | i \in \Z\};\\
\forall (\alpha, i) \in \Sigma: &\epsilon \bowtie
\alpha[\epsilon,\dots, \epsilon];\\
\forall i \in \Z: &\triangle_i[a, b] \bowtie \triangle_i[a, b]
\wedge \sqcap_i[a] \bowtie \sqcap_i[a]
\wedge \cap_i[a] \bowtie \cap_i[a];\\
\forall i, j \in \Z: i \not= j \Rightarrow
&\triangle_i[\triangle_j(a, b), \triangle_j(c, d)] \bowtie
\triangle_j[\triangle_i(a, c), \triangle_i(b, d)];\\
\forall i, j \in \Z: i \not= j \Rightarrow
&\sqcap_i[\triangle_{j + 1}(a, b)] \bowtie
\triangle_j[\sqcap_i(a), \sqcap_i(b)];\\
\forall i, j \in \Z: i \not= j \Rightarrow
&\cap_i[\triangle_{j - 1}(a, b)] \bowtie
\triangle_j[\cap_i(a), \cap_i(b)].
\end{eqnarray*}

Taking into account that some interaction systems are equivalent
in the sense that one can simulate another, and for any graph
rewriting system, there can be constructed an equivalent
interaction system, Lafont considers the simplest interaction
systems, and eventually finds an extremely simple universal
interaction system of interaction combinators~\cite{lafont97}:
\begin{eqnarray*}
&\Sigma = \{(\delta, 2), (\gamma, 2), (\epsilon, 0)\};&\\
\delta[x, y] \bowtie \delta[x, y],
&\gamma[\delta(a, b), \delta(c, d)] \bowtie
\delta[\gamma(a, c), \gamma(b, d)],
&\gamma[x, y] \bowtie \gamma[y, x];\\
\epsilon \bowtie \delta[\epsilon, \epsilon],
&\epsilon \bowtie \epsilon,
&\epsilon \bowtie \gamma[\epsilon, \epsilon].
\end{eqnarray*}

Bechet went even farther, and managed to find a universal
interaction system with only two agents~\cite{bechet}, thus even
simpler than Lafont's one.
However, the price was that the rules had to be too complicated.
He also stated a question which, as far as we know, still remains
open: is it possible to find a universal system with
${\Sigma = \{(\xi, 2), (\epsilon, 0)\}}$?
There is obviously no way to simplify this signature, however,
we can get back to graph rewriting systems, focusing on their
implementations using random access memory heap, and try to apply
methods similar to those leading to interaction combinators.

%% file: op.tex
\section{Primitive graph operations}

In order to construct the simplest automata, we first define
a set of their possible states as
$$
S(M) = \{(f_0, f_1) | f_0, f_1: M \rightarrow M\},
$$
where $M$ is a finite set.
Each element of $S(M)$ can be considered as a finite directed
graph with the set of nodes $M$, which has exactly two arrows
from each node, the arrows being labeled $0$ and $1$.
Let us take a look at the simplest operations on those graphs:
$$
e[b_0b_1 \dots b_n := a_1 \dots a_m]: S(M) \rightarrow S(M),
$$
where $e \in M$ and $\forall i: a_i, b_i \in \{0, 1\}$.
We will require
$$
e[b_0b_1 \dots b_n := a_1 \dots a_m](f_0, f_1) = (g_0, g_1)
$$
to have certain properties.
Namely, if
$$
a = f_{a_1}(\dots f_{a_m}(e) \dots),\quad
b = f_{b_1}(\dots f_{b_n}(e) \dots),
$$
$a$ must be equal to ${g_{b_0}(b)}$, and $b$ must be the only
point where ${(g_0, g_1)}$ differs from $(f_0, f_1)$:
\begin{eqnarray*}
&a = g_{b_0}(b);\\
&i \not= b_0 \Rightarrow \forall x \in M: g_i(x) = f_i(x);\\
&\forall x \in M: x \not= b \Rightarrow g_{b_0}(x) = f_{b_0}(x).
\end{eqnarray*}

We will take the liberty to illustrate these primitive graph
operations by its implementation in the C programming language:
\begin{verbatim}
struct node {
        struct node *left, *right;
} state[MEMSIZE];

void op(struct node *element)
{
        element->left->right = element->right->left->left; 
}
\end{verbatim}
Here, if every structure's fields all point to nodes in
the array itself, the state corresponds to an element of
${S(M)}$, ${|M|}$ being equal to the array size.
Then, calling the function basically maps the array from
one state to another, so it directly implements ${e[01 := 100]}$,
$e$ corresponding to the function's argument.

%% file: embed.tex
\section{Embedding logical operations}

One can notice that the graphs of the introduced type are similar
to the internal representation of S-expressions, or lists, in the
LISP programming language.
The only difference is that our graphs do not contain any atoms,
or data, thus being a pure recursive version of S-expressions.

An arbitrary graph rewriting rule defined for the elements of
${S(M)}$ will result in a graph rewriting system with only one
type of nodes, each one having exactly two outgoing arrows.
The latter makes the introduced structure relevant to the open
question by Bechet mentioned in the beginning of this paper, once
we find a composition
$$
T = e[b_0b_1\dots b_n := a_1\dots a_m] \circ \dots \circ
e[y_0y_1\dots y_q := x_1\dots x_p]
$$
that makes our system universal as its only graph rewriting rule.

But first, let us have at least conditional statements embedded.
We may use the same approach conditional statements ${B\, M\, N}$
are implemented in $\lambda$-calculus: for true
${B = \lambda x.\lambda y.x}$, the expression is $M$; for false
${B = \lambda x.\lambda y.y}$, the expression is $N$.
So, let $M$ be embedded into some subgraph starting at
${m = f_0(f_0(e))}$, $N$---at ${n = f_1(f_0(e))}$, and a boolean
value---at ${b = f_1(e)}$ so that ${f_1(f_0(b)) = f_0(b)}$
for true and ${f_1(f_0(b)) = f_1(b)}$ otherwise.
Then, a composition of primitive graph operations applied to
this graph results in a graph
${(g_0, g_1) = (e[001 := 00] \circ e[011 := 10])(f_0, f_1)}$
that has $m$ or $n$ at $g_0(g_1(g_0(g_1(e))))$ for true and
false, respectively.

Using the similar methods, the reader can easily produce
logical ``and,'' logical ``or,'' negation and so on,
producing embedded version of nearly every computable function.
Assuming $M$ to be unbounded, one can also prove
Turing-completeness of such a system, by simulating
any other universal system, like interaction combinators, or
a universal Turing machine.

Finally, let us notice that for any
${e[b_0b_1 \dots b_n := a_1 \dots a_m]}$, the graph
${(f_0, f_1)}$ is a fixed point if and only if
${f_{b_0}(f_{b_1}(\dots f_{b_n}(e) \dots)) =
f_{a_1}(\dots f_{a_m}(e) \dots)}$.
This also provides an obvious way to construct fixed points
for any composition $T$ of primitive graph operations.
And since a finite-state machine with the set of states
${S(M)}$ and transition function $T$ practically stops at
the first fixed point reached from the initial state,
this way we are able to simulate a system halt.

%% file: app.tex
\section{Possible applications}

Let encoding ${c: \Lambda \leftrightarrow S(M)}$ for
$\lambda$-expressions be compatible with $T$ in the sense that
$$
\exists n \in \mathbb{N}: c^{-1}(T^n(c(P))) = Q
\wedge T(c(Q)) = c(Q),
$$
where $P$ is a $\lambda$-expression, and $Q$ is its normal form.
Once the minimal value of $n$ is proportional to the number of
steps on optimal reduction from $P$ to $Q$ asymptotically by the
complexity of $P$, transition $T$ makes the corresponding
finite-state machine practically the simplest implementation of
optimal reduction.
Currently, our primary goal is to answer the question whether
there exists such a composition with compatible encoding.

But although we mainly pursue the simplest automata that would
implement optimal reduction of $\lambda$-expressions,
the ``blind'' rewriting systems considered in this paper might
have some other possibly useful applications as well.

With respect to composition, the primitive graph operations
defined above obviously generate a quite simple but still unusual
algebraic structure which, as far as we know, does not directly
correspond to any well-known mathematical structure.
If this is the case, one could probably be interested in
analyzing the generated structure.
Otherwise, an attempt to find such a correspondence might deserve
a separate research.

We believe that the ideas described in this paper could be
interesting from the view point of actual implementing graph
reduction for functional programming languages.
Moreover, simplicity and low-level transparency of the structures
discussed above might possibly make these ideas a fruitful
direction for computer hardware significantly differing
from the existing.

One can also consider input/output techniques for such a system,
for instance, using some ideas behind TTA, transport triggered
architecture.
Namely, output can be implemented by introducing
side effects of accessing particular nodes within its memory.
In turn, input may be done by transition against $i \not= e$
initiated from outside.